\documentclass[fleqn,10pt]{wlscirep}
\usepackage[T1]{fontenc}
\usepackage{graphicx}
\usepackage{dcolumn}
\usepackage{bm}
\usepackage{mathrsfs}
\usepackage{hyperref}
\usepackage[mathlines]{lineno}
\usepackage[utf8]{inputenc}
\usepackage{textcomp}

\title{Dual-mode superconducting diode effect enabled by in-plane and out-of-plane magnetic field }

\author[1,3,*,\dag]{Chengyu Yan}
\author[1,\dag]{Huai Guan}
\author[4,5]{Zhenyu Zhang}
\author[4,5]{Yiheng Sun}
\author[1]{Qiao Chen}
\author[1]{Xinming Zhao}
\author[2]{Chuanwen Zhao}
\author[5,*]{James Jun He}
\author[1,3,*]{Shun Wang}

\affil[1]{National Gravitation Laboratory, MOE Key Laboratory of Fundamental Physical Quantities Measurement, and School of Physics, Huazhong University of Science and Technology, Wuhan 430074, China}
\affil[2]{State Key Laboratory of Magnetic Resonance and Atomic and Molecular Physics, National Center for Magnetic Resonance in Wuhan, Wuhan Institute of Physics and Mathematics, Innovation Academy for Precision Measurement Science and Technology, Chinese Academy of Sciences, Wuhan 430071, Hubei, China}
\affil[3]{Institute for Quantum Science and Engineering, Huazhong University of Science and Technology, Wuhan 430074, China}
\affil[4]{International Center for Quantum Design of Functional Materials (ICQD), 
		Hefei National Research Center for Interdisciplinary Sciences at the Microscale, 
		University of Science and Technology of China, Hefei, Anhui 230026, China}
\affil[5]{Hefei National Laboratory, Hefei, Anhui 230088, China	}

\affil[$\dag$]{C. Yan and H. Guan contribute equally to the work. }
\affil[*]{corresponding author(s): Chengyu Yan(chengyu$\_$yan@hust.edu.cn), James Jun He(jun$\_$he@ustc.edu.cn), Shun Wang(shun@hust.edu.cn)}

\begin{abstract}
The discovery of the superconducting diode effect (SDE) has been cherished as a milestone in developing superconducting electronics. Tremendous efforts are being dedicated to realizing SDE in a wide variety of material platforms. Despite the diversity in the hosting materials and device designs, SDE is usually operated in a single mode which is enabled by either out-of-plane or in-plane magnetic field/magnetization. In this work, we report the realization of a dual-mode SDE in 2H-$\mathrm{NbS_2}$/2H-$\mathrm{NbSe_2}$ heterostructures where both the out-of-plane magnetic field $B_{\perp}$ and in-plane magnetic field $B_{||}$ can independently generate and manipulate SDE. The two modes share similar diode efficiency but differ in two aspects: 1. $B_{\perp}$-induced SDE is activated by a field on the order of 1 mT while $B_{||}$-induced SDE requires a field on the order of 100 mT; 2. $\eta$ of $B_{\perp}$-induced SDE exhibits a square-root like temperature dependence while $\eta$ of $B_{||}$-induced SDE takes a more linear-like one. We demonstrate that the dual-mode SDE is most likely a result of mirror symmetry breaking along multiple orientations. Thanks to the two orders difference in the operational field for the two modes, we propose a dual-functionality device scheme to showcase the potential of the dual-mode SDE in realizing advanced superconducting architecture, where fast polarity-switching functionality is implemented with $B_{\perp}$-induced SDE and high-fidelity functionality is enabled with $B_{\perp}$-induced SDE.     
\end{abstract}

\begin{document}                              
\flushbottom
\maketitle

\thispagestyle{empty}


 
Symmetry-breaking has stimulated some of the most celebrated hallmarks in physics\cite{DHC21}. For instance, semiconductor diodes harness nonreciprocal conductance due to the lifting of inversion symmetry and lay the foundation of modern electronic systems. In the superconducting realm, symmetry-breaking is also a driving force in the realization of unconventional superconductivity\cite{SKK19,MDY22, WZY23,ZGY24} such as Fulde–Ferrell–Larkin–Ovchinnikov state (FFLO) and orbital FFLO state or unique superconducting functionality as exemplified by superconducting diode effect (SDE)\cite{AML20,HNC23,LPX24,YNF22,HTN22,ISB22,DIY22,NFW23,JKH22}. The superconducting diode effect, signified by the flow of dissipationless supercurrent in one direction and dissipative normal current in the opposite direction, has been extensively studied in a wide range of material platforms for both superconducting films and Josephson junctions\cite{AML20,HNC23,LPX24,BBF22,BFC22,GPB24,WWX22,PCS22,JKY22,SVL23}. The dissipationless nature makes the superconducting diode a promising candidate for superconducting electronics\cite{BRA19} in both classical and quantum regimes. Regardless of the details in the device design, it is now commonly agreed that inversion and time-reversal symmetry should be simultaneously lifted to enable SDE\cite{NFW23}. There are rather diverse approaches to breaking the inversion symmetry in a superconducting system, such as stacking different materials\cite{AML20,HNC23,YYL19}, layer twisting\cite{DPZ21,YDJ21,DDY23} and nano-patterning\cite{LJW21,GTK22}. In comparison, time-reversal symmetry is routinely lifted by a magnetic field or magnetization, apart from several reports on zero-field SDE whose underlying mechanisms are not yet clear\cite{WWX22,LPX24}.

For the bulk of previous works, it is noteworthy that SDE can only be activated by a magnetic field or magnetization along a critical orientation, either in-plane or out-of-plane, while it vanishes for the orthogonal orientations\cite{NFW23}. These realizations are thus referred to as single-mode SDE in the rest of the manuscript. The critical orientation is closely associated with the inversion or mirror symmetry breaking orientation\cite{NFW23}. Note that the critical orientation for SDE arising from the Meissner screening effect\cite{SVL23} and vortex dynamics\cite{GPB24,GTK22,FYJ25} may not follow the same criteria. It is then natural to expect that both in-plane and out-of-plane magnetic field/magnetization can independently induce and manipulate SDE in a system where inversion or mirror symmetry breaking occurs along multiple orientations. As a result, the SDE can operate in two modes (hereafter denoted as dual-mode SDE). Hints in favor of this hypothesis have emerged in several material platforms. SDE triggered by an in-plane or out-of-plane magnetic field, but not both, has been reported in twisted graphene\cite{YDJ21,DDY23}. In few-layer $\mathrm{MoTe_2}$\cite{DCZ24}, it has been demonstrated that an out-of-plane magnetic field activates SDE in deep superconducting regime while an in-plane magnetic field enables moderate magnetochiral anisotropy in the superconducting transition regime. In addition, it is seen in few-layer $\mathrm{NbSe_2}$ that an out-of-plane magnetic field can result in SDE meanwhile an in-plane field can modulate diode efficiency \textit{iff} in the presence of out-of-plane field\cite{BBF22}, a complementary behavior is reported in an Al/InAs/Al junction\cite{BFC22}. However, well-established evidence for dual-mode SDE is lacking\cite{HNC23}.

\begin{figure}
	\centering
	\includegraphics[width=0.8\textwidth]{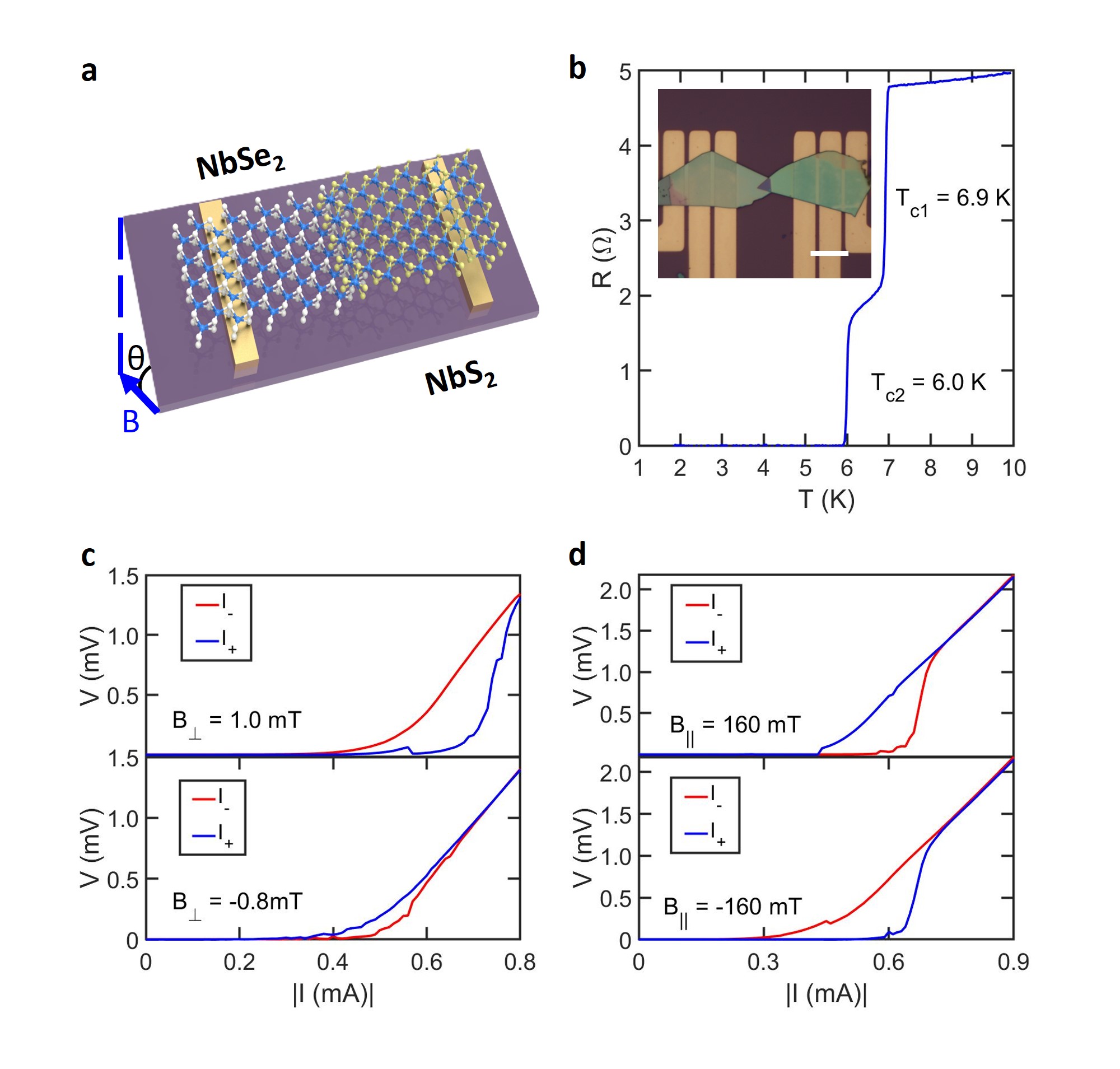}
	\caption{Experiment setup and device characterization. \textbf{a}, The device is made of 2H-$\mathrm{NbS_2}$/2H-$\mathrm{NbSe_2}$ heterostructure. The 2H-$\mathrm{NbS_2}$ flake hangs on the top, while the 2H-$\mathrm{NbSe_2}$ flake locates at the bottom. The thickness of the flakes spans from 25 to 45 nm. The entire device is mounted on a rotatable sample holder whose polar angle $\theta$ can be continuously tuned from -90$^\circ$ to 90$^\circ$. At $\theta=\pm 90^\circ$, the magnetic field is perpendicular to the flake plane. At $\theta=0^\circ$, the magnetic field is in-plane and along the long edge of the electrode. \textbf{b}, Temperature dependence of 4-terminal resistance. The sharp drops at $T_{c1}$ = 6.9 K and $T_{c2}$ = 6.0 K correspond to the superconducting transition of $\mathrm{NbSe_2}$ and $\mathrm{NbS_2}$ flake, respectively. Inset shows an optical image of the device. The scale bar is 10 $\mu$m. \textbf{c\&d}, Nonreciprocity of critical current $I_{c}$ in the presence of out-of-plane ($\theta= 90^\circ$) and in-plane ($\theta= 0^\circ$) magnetic field, respectively. Note that the critical current is recorded at the forward sweeping, i.e., from 0 to a given value, in the work unless otherwise specified. Results of device 2, which has a steep switching in IV characteristic and shows qualitatively similar results in SDE,  are appended in Extended Data Fig.\ref{ExtFig3}-\ref{ExtFig6}.         }
	\label{1}
\end{figure}

In this work, we report the realization of a dual-mode SDE in 2H-$\mathrm{NbS_2}$/2H-$\mathrm{NbSe_2}$ heterostructure, where both the out-of-plane magnetic field $B_{\perp}$ and in-plane magnetic field $B_{||}$ can independently generate and manipulate SDE. This specific combination of the hosting materials not only enables a pronounced Josephson coupling and sizable critical current but also breaks inversion or mirror symmetry along multiple orientations. The dual-mode SDE is unambiguously confirmed by the coexistence of $B_{\perp}$- and $B_{||}$-induced SDE when the magnetic field points to an arbitrary polar angle, differing from 0$^\circ$ or $\pm90^\circ$, with respect to the flake plane. The maximum diode efficiency obtained in both modes can reach $\sim12\%$. It is particularly interesting to mark that the SDE is activated by $B_{\perp}$ on the order of $\sim$1 mT while it is enabled by $B_{||}$ on the order of $\sim$100 mT. Meanwhile, the temperature dependence of  $B_{\perp}$-induced SDE is noticeably different from that of $B_{||}$-induced SDE. We believe the reduction from $D_{3h}$ to $C_3$ or $C_{3v}$  symmetry due to the formation of 2H-TMDC heterostructure is likely to play a vital role in generating the observed dual-mode SDE. Regarding device implication, the dual-mode SDE may enable complicated functionality that is not feasible for its single-mode counterpart. $B_{\perp}$-induced SDE only requires a tiny operational magnetic field, hence it can be readily integrated with an on-chip nanomagnet to realize fast polarity-switching functionality. On the other hand,  $B_{||}$-induced SDE needs a relatively large magnetic field, therefore making it insensitive to local magnetic fluctuation in a logical circuit and suitable for high-fidelity operation. We believe our results not only enrich the design of SDE by incorporating multiple control knobs into a single platform but also unleash its potential in implementing more advanced superconducting electronic architecture.

\section*{Observation of $B_{\perp}$- and $B_{||}$-induced SDE }

We adopted a modified dry transfer protocol to fabricate $\mathrm{NbS_2}$/$\mathrm{NbSe_2}$ heterostructure to mitigate degradation of $\mathrm{NbS_2}$ flake\cite{ZYC22}. Energy dispersive spectrometer (EDS) analysis enclosed in  Fig.S1 of the supplementary information indicates that the heterostructure, especially the interface, is almost free of oxygen. It is important to stress that $\mathrm{NbS_2}$ and $\mathrm{NbSe_2}$ flakes with comparable thickness were stacked together to form the heterostructure, such that the two flakes were equipped with similar superconducting order parameters and other electronic properties. This enabled a pronounced Josephson coupling between the flakes and a sizable critical current. The resultant Josephson junctions were characterized by a standard 4-probe setup in a rotatable sample holder, whose relative polar angle $\theta$ with respect to the magnetic field could be tuned continuously between $\pm 90^\circ$, as illustrated in Fig.\ref{1}a. At $\theta=\pm 90^\circ$, the magnetic field is perpendicular to the flake plane. On the other hand, the field is in-plane and along the long edge of the electrodes at $\theta=0^\circ$. We present the results rendered from device 1 in the main text and append results of device 2, showing qualitatively similar behavior in SDE, in Extended Data Fig.\ref{ExtFig3}-\ref{ExtFig6}.

\begin{figure}
	\centering
	\includegraphics[width=0.8\textwidth]{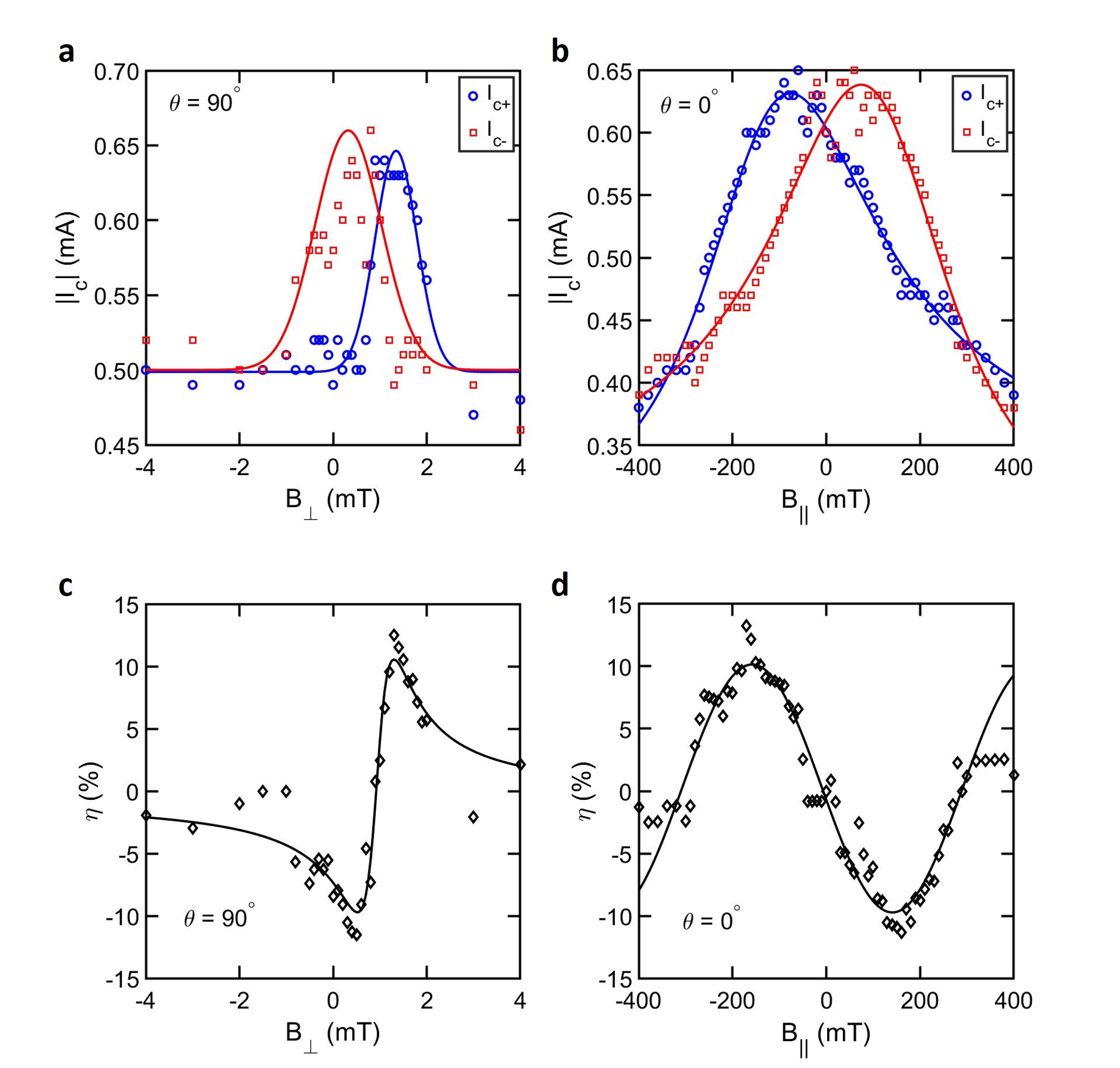}
	\caption{Evolution of superconducting diode effect as a function of magnetic field. \textbf{a\&c}, Critical current in both current bias branches and diode efficiency $\eta$ with respect to out-of-plane magnetic field ($\theta= 90^\circ$). \textbf{b\&d}, Critical current and $\eta$ with respect to in-plane magnetic field ($\theta= 0^\circ$). The solid curves in the plots serve as guide to the eye.  }
	\label{2}
\end{figure}

Upon cooling down, two superconducting transitions could be identified, as shown in Fig.\ref{1}b. The one at 6.9 K corresponded to superconducting transition of 2H-$\mathrm{NbSe_2}$ flake\cite{YKS19}, the other one at 6.0 K matched that of 2H-$\mathrm{NbS_2}$ flake\cite{XWZ16} and the junction. The sharp transition highlights the high quality of the heterostructure. The superconducting diode effect can be revealed from the current-voltage characteristics. Note that all the current-voltage characteristics were recorded at the forward sweeping, i.e., from 0 to a given value, in the work unless otherwise specified. It is then interesting to recognize that both out-of-plane magnetic field $B_{\perp}$ and in-plane magnetic field $B_{||}$ could trigger pronounced superconducting diode effect manifested as the nonreciprocity between the critical current $I_c$ of positive current branch $I_{+}$ and negative one $I_{-}$, as highlighted in Fig.\ref{1}c and d, respectively. The sign of nonreciprocity in $I_c$ was reversed with the reversal of magnetic field orientation. We have carefully ruled out that the superconducting diode effect may arise from the fluctuation in $I_c$ by repeating the measurements (Extended Data Fig.\ref{ExtFig1}a). Also, the superconducting diode effect exhibited long-term stability in half-wave rectification regardless of the orientation of the magnetic field (Extended Data Fig.\ref{ExtFig1}b).

The superconducting diode effect of the $\mathrm{NbS_2}$/$\mathrm{NbSe_2}$ Josephson junction is highly sensitive to the applied magnetic field. $I_{c+}$ and $I_{c-}$ follow Gaussian-like distribution with respect to $B_{\perp}$ as depicted in Fig.\ref{2}a, peaking at 1.3 mT and 0.5 mT, respectively. The distribution of $I_{c+}$ and $I_{c-}$, in turn, lead to a skewed lineshape of superconducting diode efficiency $\eta=\frac{I_{c+}-|I_{c-}|}{I_{c+}+|I_{c-}|}$ as a function of $B_{\perp}$, with the two antinodes locating at 1.3 mT and 0.5 mT, as enclosed in Fig.\ref{2}c. The maximum $\eta$ is $\sim12\%$. It is also realized that $\eta$ does not vanish at $B_{\perp}=0$ but at $B_{\perp}=0.9$ mT. However, we would like to stress that the nonvanishing $\eta$ at zero field is more likely due to uncalibrated residual magnetic field in the circuitry or flux trapping\cite{ZWW23} during a given thermal cycling process, since the drift of $\eta$=0 node does not show consistent behavior in different runs of thermal cycling. If the drift is manually calibrated out, the antinodes would be subsequently corrected to $\pm$0.4 mT which is in line with the expectation of field-driven SDE\cite{YNF22,NFW23}. Then, we move to the effect of the in-plane magnetic field as presented in Fig.\ref{2}b and d. The distribution of $I_{c+}$ and $I_{c-}$ mirrors each other with respect to $B_{||}=0$, reaching maximum at $\pm160$ mT. The resultant $\eta$ can be nicely captured by a sinusoidal function between $B_{||}=\pm300$ mT and then seemingly saturates in the larger field. The maximum $\eta$ again can reach $\sim12\%$. Now that $\eta$ in the presence of $B_{\perp}$ and $B_{||}$ exhibits distinct lineshape, it is perhaps helpful to comment that the lineshape has been proposed to be associated with the underlying mechanism of the SDE such as if there are higher order terms in the current phase relation\cite{MSL22}. However, there seems to be no consensus on the argument of lineshape for the time being.    

It is necessary to stress that dual-mode SDE is absent in homostructure. In $\mathrm{NbSe_2}$/$\mathrm{NbSe_2}$ homostructure device 3, we have only managed to observe a rather weak $B_{||}$-induced SDE ($\eta\sim3\%$) which is perhaps due to unintended twist between the layers, as enclosed in Extended Data Fig.\ref{ExtFig7}. In $\mathrm{NbSe_2}$/$\mathrm{NbSe_2}$ homostructure device 4, SDE is always negligible regardless of the orientation of the magnetic field. The dimensions of the heterostructures and homostructures are summarized in the Methods.    

\section*{Coexistence of $B_{\perp}$- and $B_{||}$-induced SDE at arbitrary $\theta$ }

\begin{figure}
	\centering
	\includegraphics[width=0.8\textwidth]{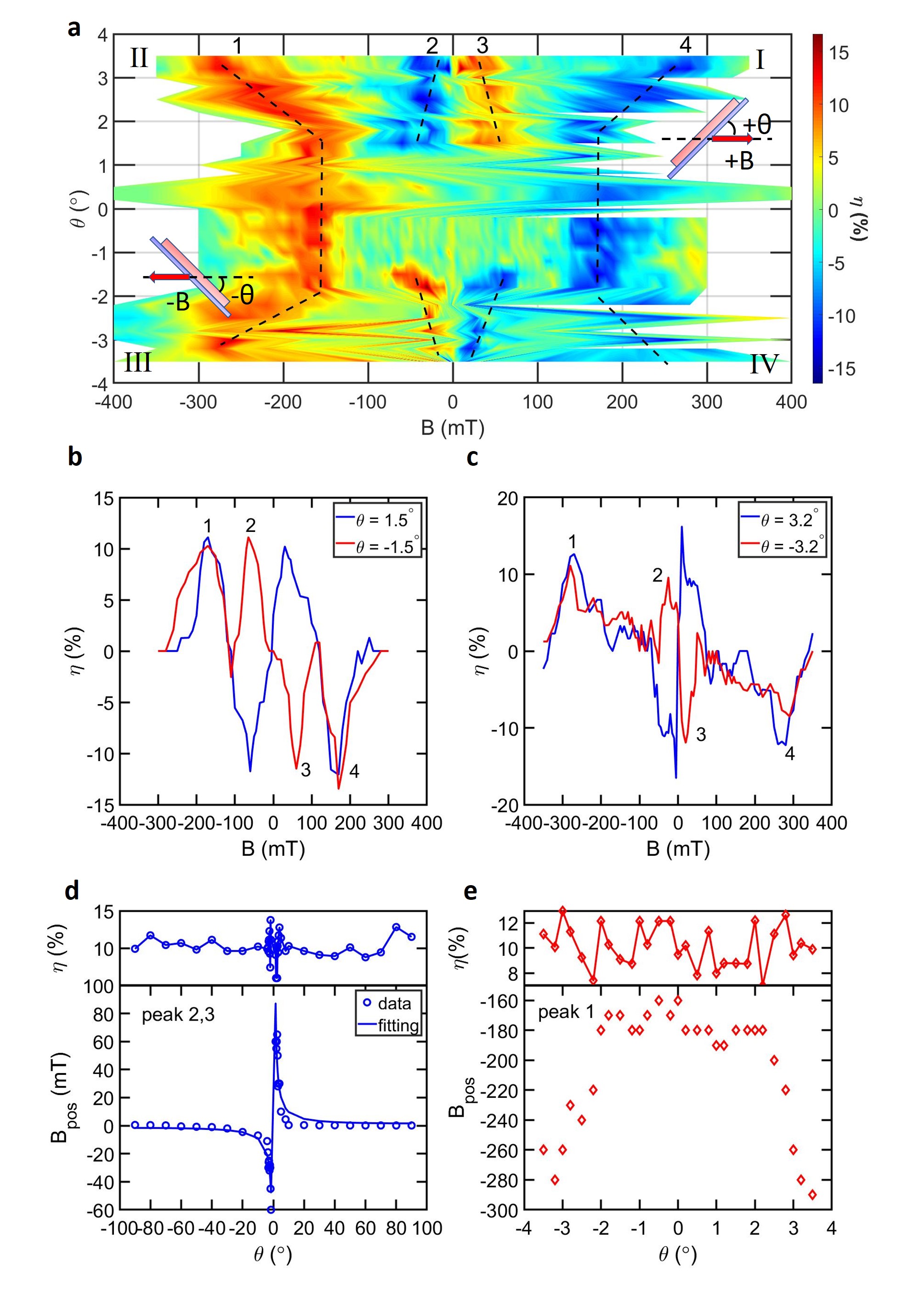}
	\caption{Superconducting diode effect at arbitrary $\theta$. In the experiment, the polar angle $\theta$ of flake plane is modulated between $\pm90^\circ$ while the orientation of the magnetic field is fixed. \textbf{a}, SDE within small angle limit. 4 SDE peaks 1-4, marked by the dash curves, can be recognized. Peaks 1$\&$4 are dominated by in-plane component of the magnetic field, whereas peaks 2$\&$3 are dominated by out-of-plane component. The insets in sector I and III of the colorplot illustrate the relative field orientation. \textbf{b\&c}, linecut at $\theta=\pm1.5^\circ$ and $\theta=\pm3.2^\circ$, respectively. \textbf{d}, diode efficiency $\eta$ and peak position $B_{pos}$ of $B_{\perp}$-induced SDE peaks. $B_{pos}$ follows a $\mathrm{1/sin\theta}$ trend as indicated by the solid fitting curve. Representative SDE data at large $\theta$ can be found in Extended Data Fig.\ref{ExtFig2}. Note that, we stick to the $\eta>0$ branch in the plot, $B_{pos}$ at $\theta>0$ corresponds to the position of peak 3 in sector I while $B_{pos}$ at $\theta<0$ corresponds to that of peak 2 in sector III. \textbf{e}, $\eta$ and $B_{pos}$ of $B_{||}$-induced SDE peak 1.               }
	\label{3}
\end{figure}

Since it only requires a $B_{\perp}$ on the order of 1 mT to activate SDE while a $B_{||}$ of $\sim$100 mT is necessary to trigger SDE, it is likely to speculate if $B_{||}$-induced SDE arises from nothing but inadequate calibration of polar angle $\theta$\cite{HNC23}. For instance, at a total magnetic field of 100 mT, a relatively small misalignment of $\theta=0.01$ rad (or 0.5$^\circ$) can already lead to a residual $B_{\perp}$ component of 1 mT. This possibility can be ruled out in our experiment. First, $\theta$ can be calibrated to 0.1$^\circ$ as detailed in Extended Data Fig.\ref{ExtFig2}a. Second and more conclusive, we have observed the coexistence of $B_{\perp}-$ and $B_{||}-$induced SDE at $\theta$ between $\pm3.5^\circ$. 

Before jumping to SDE at arbitrary $\theta$, it is worthy of commenting on the experiment setup a bit more. At a given $\theta$, reversing the polarity of the total magnetic field B will no doubt reverse the polarity of its out-of-plane component $B_{\perp}$ and in-plane component $B_{||}$ simultaneously. At a given B, swapping the sign of $\theta$ will only alter the polarity of $B_{\perp}$ but preserve the polarity of $B_{||}$ in our setup, see Fig.\ref{1} or inset of Fig.\ref{3}a. This allows us to carry out a comprehensive polarity analysis of SDE.

With the clarification of the experiment setup, then we can examine the experimental results. When gradually reducing $\theta$ from $\pm 90^\circ$ to $0^\circ$, there is a noticeable change in the number of antinodes in $\eta$-B relation. It starts with two antinodes as shown in Fig.\ref{2}c, then 4 antinodes happen at $1.0^\circ<\theta<3.5^\circ$ or $-3.5^\circ<\theta<-1.0^\circ$ (Fig.\ref{3}), and eventually two antinodes behavior restores as presented in Fig.\ref{2}d. This evolution differentiates the multiple antinodes in our work from that in previous reports on sign reversal of SDE\cite{PCS22,CBR23,SVL23}. Multiple antinodes behavior in previous works\cite{PCS22,CBR23,SVL23} can be driven by pure $B_{\perp}$ or $B_{||}$. In contrast, the multiple antinodes behavior is absent at both $\theta=\pm90^\circ$ and $0^\circ$ in our heterostructure. A colorplot highlighting the multiple antinodes behavior is presented as Fig.\ref{3}a, with typical linecut at $\pm1.5^\circ$ and $\pm3.2^\circ$ appended in Fig.\ref{3}b and c, respectively. Here SDE peaks 1-4 can be readily recognized. Peak 1 and peak 4, occurring in a larger magnetic field, are time-reversal pairs since they are antisymmetric 
with respect to B = 0. Similarly, peaks 2 and 3 emerging in a smaller field are time-reversal pairs. Also, it is particularly interesting to note that the polarity of peak 1 or 4 is insensitive to the sign of $\theta$, while the polarity of peak 2 or 3 depends on the sign of $\theta$. We attribute peaks 1 and 4 to $B_{||}$-induced SDE while peaks 2 and 3 to $B_{\perp}$-induced SDE by analyzing the correlation between the polarity of SDE peaks and polarity of $B_{||}$ or $B_{\perp}$ component in the four sectors of the colorplot.  
 
First, let us focus on peak 1 in sector II ($B_{||}<0$, $B_{\perp}>0$) and sector III ($B_{||}<0$, $B_{\perp}<0$) of the colorplot. Here polarity of $B_{||}$ is fixed, polarity reversal of $B_{\perp}$ has no impact on the polarity of peak 1. On the contrary, altering the polarity of $B_{||}$ while preserving the polarity of $B_{\perp}$ can result in the polarity swap of peak 2,  as unveiled by crosschecking peak 2 in sector II ($B_{||}<0$, $B_{\perp}>0$) and its time-reversal pair peak 4 in sector IV ($B_{||}>0$, $B_{\perp}>0$). Hence, peaks 1 and 4 arise from $B_{||}$-induced SDE. Peaks 1 and 4 remain persistent at $\theta=0^\circ$.   

Similarly, examining peak 2 in sector II ($B_{||}<0$, $B_{\perp}>0$) and sector III ($B_{||}<0$, $B_{\perp}<0$) reveals that polarity reversal of $B_{\perp}$ alone can lead to the polarity change of peak 2. Then, we can also crosscheck peak 2 in sector II ($B_{||}<0$, $B_{\perp}>0$) and its time-reversal pair peak 3 in sector IV ($B_{||}>0$, $B_{\perp}>0$). It is then concluded that the polarity of peak 2 would remain the same as long as there is no change in polarity of $B_{\perp}$, regardless of the polarity of $B_{||}$. Hence, peaks 2 and 3 correspond to $B_{\perp}$-induced SDE. Peaks 2 and 3 are absent at $\theta=0^\circ$.

Therefore, we can justify that the observed $B_{\perp}$- and $B_{||}$-induced SDE do not originate from the misalignment of field orientation but have an intrinsic reason. To gain more insights into the underlying mechanisms of the two SDE modes, we carry out an assessment of the evolution of the diode efficiency $\eta$ and peak position $B_{pos}$. 

For $B_{\perp}$-induced SDE peaks, we assemble $\eta$ and $B_{pos}$ of peak 3 in sector I and peak 2 in sector III so that $\eta$ is always positive. The results are plotted in Fig.\ref{3}d. Presenting data in other ways does not alter the central conclusion. It is observed that $\eta$ of $B_{\perp}$-induced SDE peaks fluctuate around 10$\%$ throughout the entire $\pm90^\circ$ and $B_{pos}$ follows a $\mathrm{1/sin\theta}$ behavior. The difference between the data and fitting of  $B_{pos}$ mainly arises from the drift of $\eta=0$ node. These observations have three insightful implications: 1. $B_{\perp}$-induced SDE peaks may shift to an ultra-large magnetic field at $\theta\sim0^\circ$, and hence it is not observed in the experiment; 2. $B_{\perp}=\mathrm{B_{pos}sin\theta}$ remains the same for peaks 2 and 3 with respect to $\theta$, whereas $B_{||}$ changes significantly; 3. The change in $B_{||}$ has little impact on $\eta$ of peaks 2 and 3, which is noticeably different from SDE in few-layer $\mathrm{NbSe_2}$ flake where $B_{||}$ can modulate $\eta$ \textit{iff} in the presence of $B_{\perp}$\cite{BBF22}. 

For $B_{||}$-induced SDE peaks, we depict $\eta$ and $B_{pos}$ of peak 1 in Fig.\ref{3}e. Again, $\eta$ fluctuates around $10\%$. For $B_{pos}$ of peak 1, it almost remains the same within $\pm2^\circ$ and then booms by 70$\%$ with $\theta$ increasing by another $1.5^\circ$. Qualitatively speaking, the increase in $B_{pos}$ at a larger angle is not surprising in terms of in-plane projection. However, the change in $B_{pos}$ is much faster than $\mathrm{B_{pos}(0^\circ)/cos\theta}$, considering $\mathrm{cos\theta}$ changes by less than $1\%$ within $\pm4^\circ$. As a result, $B_{||}=\mathrm{B_{pos}cos\theta}$ does not remain the same but increases rapidly for $B_{||}$-induced peak 1 or 4, meanwhile $B_{\perp}$ also increases. The observations about peak 1 or 4 suggest: 1. $B_{||}$-induced SDE peaks should move to a large magnetic field at large $\theta$, where the suppression of $I_{c}$ may prevent the observation of SDE; 2. Both $B_{||}$ and $B_{\perp}$ increase for peak 1 or 4 at larger $\theta$; 3. $\eta$ is not sensitive to the enlarging $B_{||}$ and $B_{\perp}$ for peak 1 or 4, at least within a certain range. A summary of these features can be found in Table.\ref{table1}.

\begin{table*}[h]
\centering
\begin{tabular}{|p{20mm}|p{10mm}|p{15mm}|p{35mm}|p{35mm}|p{10mm}|} 
 \hline
SDE peaks & $|\theta|$ & $|B_{pos}(\theta)|$ &$|B_{\perp}(\theta)=B_{pos}(\theta)cos\theta|$ & $|B_{||}(\theta)=B_{pos}(\theta)sin\theta|$ & $\eta(\theta)$ \\ 
 \hline
 peak 1, 4 & $\Uparrow$ & $\Uparrow$ & $\Uparrow$ & $\Uparrow$ & $\Rightarrow$\\ 
 \hline
 peak 2, 3 & $\Uparrow$ & $\Downarrow$ & $\Rightarrow$ & $\Downarrow$ & $\Rightarrow$\\
\hline
\end{tabular}
\caption{Summary of the diode efficiency $\eta$ and peak position $B_{pos}$ for SDE peaks 1-4 with respect to the polar angle $\theta$. $\Uparrow$ marks the increase of the quantity, $\Downarrow$ marks the reduction, $\Rightarrow$ indicates there is no change.  }
\label{table1}
\end{table*}

\begin{figure}
	\centering
	\includegraphics[width=0.8\textwidth]{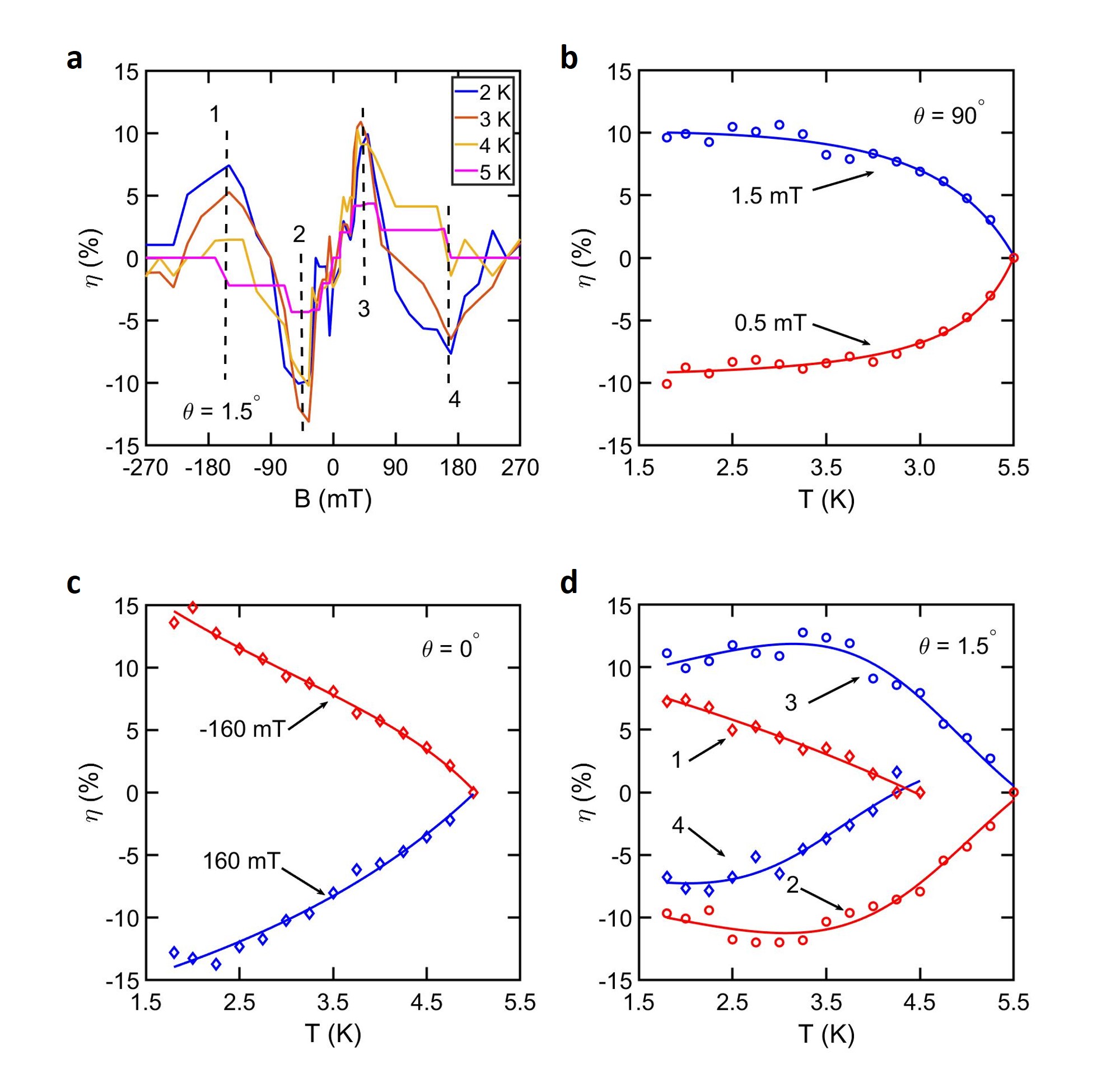}
	\caption{Temperature dependence of superconducting diode effect. \textbf{a}, $\eta$ as a function of magnetic with increasing temperature. \textbf{b} to \textbf{d} show $\eta$ as a function of temperature at $\theta=90^\circ$, $\theta=0^\circ$ and $\theta=1.5^\circ$. B = 1.5 mT and B = 0.5 mT in plot \textbf{b} correspond to the two maxima in Fig.\ref{2}c. Index 1-4 in \textbf{c} refer to SDE peaks 1-4 in Fig.\ref{4}a. Solid curves in the plots are guide to the eye.         }
	\label{4}
\end{figure}

Last but not least, we report the temperature dependence of SDE. Results obtained with $\theta$ fixed at $90^\circ$, $0^\circ$ and $1.5^\circ$ are included in Fig.\ref{4}. As demonstrated by the representative data measured at $\theta=1.5^\circ$, the position $B_{pos}$ of SDE peaks shows no explicit response to temperature change, whereas $\eta$ itself decays with increasing temperature. The robustness in $B_{pos}$ suggests that the observed SDE should not arise from self-field effect\cite{ZYC22} or Meissner screening effect\cite{SVL23}, since $B_{pos}$ in these two scenarios would be modulated by $I_c$ and therefore depends on temperature. The amplitude of $\eta$ deserves a more quantitative analysis. For $B_{\perp}$-induced SDE, see Fig.\ref{4}b and peak 2 or 3 in Fig.\ref{4}d, $\eta$ almost saturates in the low-temperature regime up to 3.5 K and then starts vanishing. It is interesting to note that $\eta \propto \sqrt{1-\frac{T}{T_c}}$ at $\theta=90^\circ$ (Fig.\ref{4}a), in line with generalized Ginzburg–Landau theories that concerns the third order term of finite-pairing momentum\cite{HTN22,DIY22}. On the contrary, $\eta$ of $B_{||}$-induced SDE, see Fig.\ref{4}c and peak 1 or 4 in Fig.\ref{4}d, exhibiting a more linear-like temperature dependence. Previous experimental and theoretical studies unveils quite diverse temperature dependence of SDE\cite{NFW23,MZL25}, possibly due to the variations in device fabrication or disorder strength\cite{NFW23}. In our device, the two SDE modes are observed in the same device but still undertake different temperature dependence, hence it provides an unprecedented opportunity to understand temperature of SDE in general.

\section*{Discussion}
\begin{figure}
	\centering
	\includegraphics[width=0.8\textwidth]{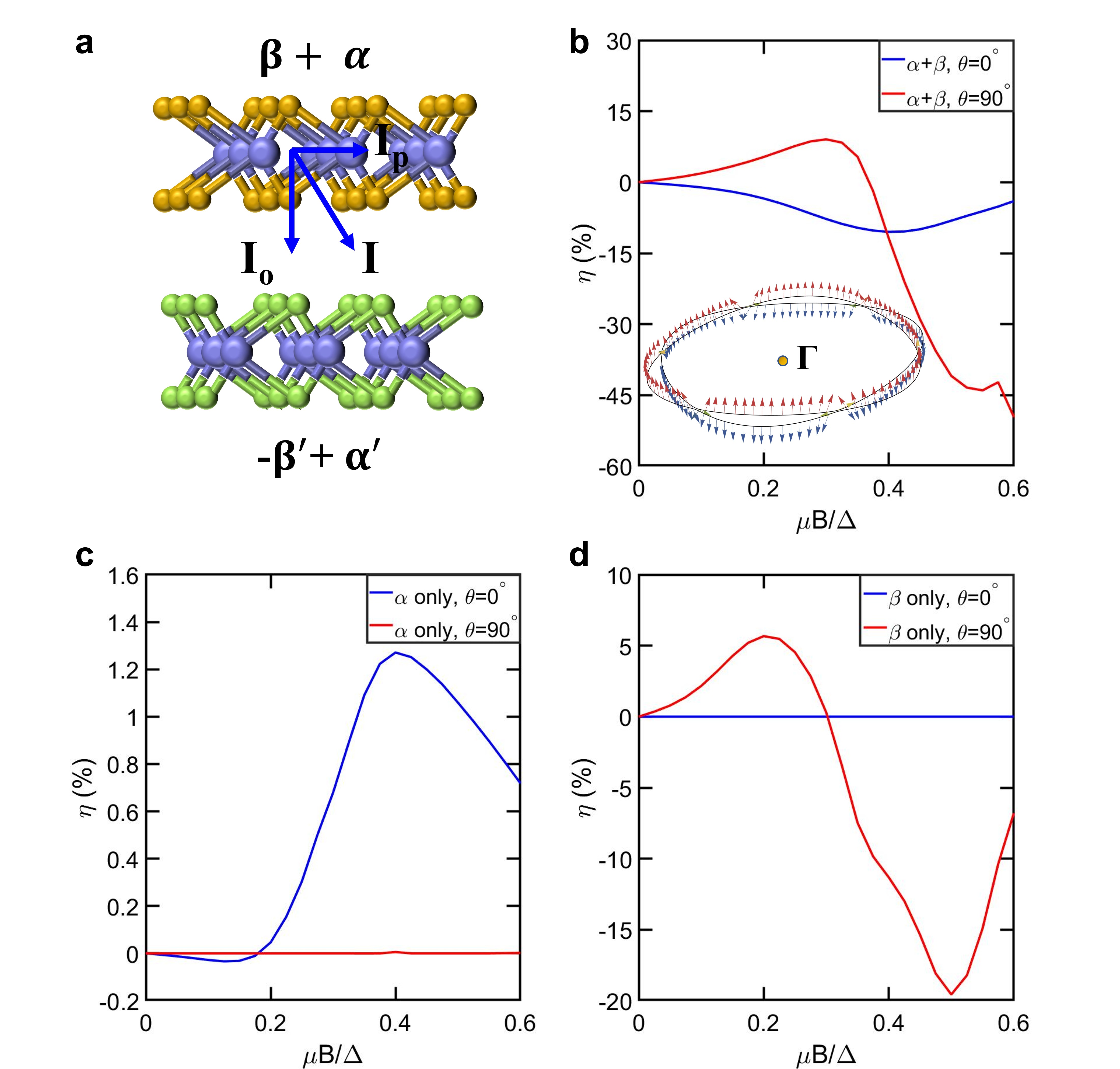}
	\caption{Possible underlying mechanisms of the dual-mode SDE. \textbf{a}, an illustration of the modeled system. Both the Josephson junction and the flakes contributes to the diode effect. To bring the contribution of the flake into account and meanwhile keep the modeling computational affordable, we consider a inclined interlayer current between two  infinitely large monolayers. The upper layer is with a total spin orbit coupling (SOC) of $\beta+\alpha$, where $\beta$ denotes the strength of Ising SOC and $\alpha$ represents the strength of Rashba SOC. The lower layer is with a SOC of $-\beta^\prime+\alpha^\prime$. The parameters can be found in the supplemental materials. \textbf{b} to \textbf{d}, the calculated diode efficiency in the presence of both SOC, Rashba SOC only and Ising SOC only as a function of normalized magnetic field $\mu B/\Delta$, respectively. $\mu$ is the Bohr bohr magneton, $\Delta$ is the order parameter. The insets in plot \textbf{b} shows the spin textures on the Fermi surfaces near the $\Gamma$ point.  
      }
	\label{5}
\end{figure}

\begin{figure}
	\centering
	\includegraphics[width=0.8\textwidth]{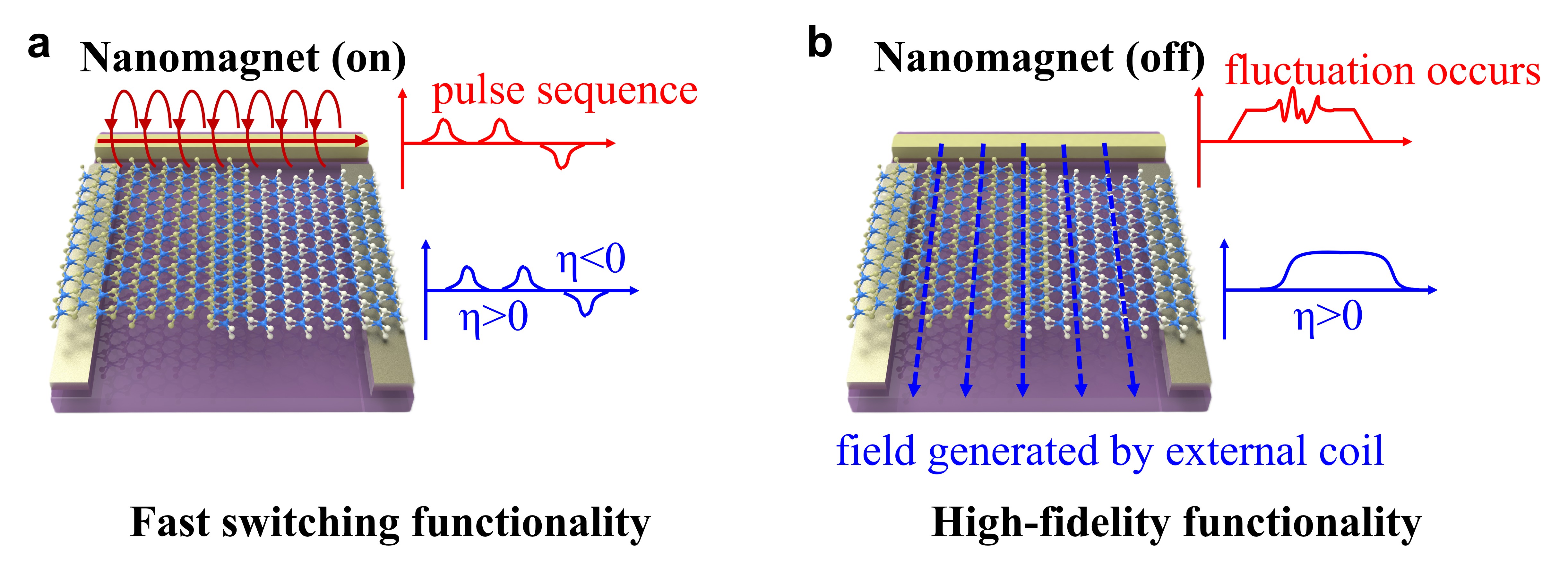}
	\caption{Potential application of dual-mode SDE in realizing advanced superconducting architecture. \textbf{a}, using an on-chip nanomagnet can drive a fast polarity switching with $B_{\perp}$-induced SDE. \textbf{b}, when the device is operated in the $B_{||}$-induced SDE mode, it requires a magnetic field that is much larger than the local magnetic field fluctuation in an integrated circuit, and thus makes it suitable for high-fidelity polarity holding or flipping operation. 
    }
	\label{6}
\end{figure}

The superconducting diode effect has been realized in a wide variety of material platforms. However, the majority of the previous works focus on single-mode SDE enabled by out-of-plane or in-plane magnetic field/magnetization, as summarized in several review papers\cite{NFW23,MZL25}. Here, we have observed both $B_{\perp}$- and $B_{||}$-induced SDE and demonstrate that they behave differently in several aspects. It is then highly desirable to explain the occurrence of these two types of SDE and their discrepancies.

It is instrumental to start the analysis of the underlying mechanisms in terms of symmetry. Stacking  2H-$\mathrm{NbSe_2}$ with 2H-$\mathrm{NbS_2}$ will reduce the symmetry from $D_{3h}$ into $C_3$ or $C_{3v}$, especially considering the few layers in the vicinity of the interface will dominate the property of the junction. At first glance, the argument based on symmetry indeed implies that both Ising and Rashba spin-orbit coupling (SOC) can emerge and contribute to $B_{\perp}$- and $B_{||}$-induced SDE\cite{BAL25}, respectively. More interestingly, the Ising SOC is usually an order stronger than Rashba SOC in TMDC materials\cite{JOI15}, naturally suggesting a much smaller $B_{\perp}$ is required to induce the same finite-pairing momentum compared to its $B_{||}$ counterpart. It seems that the naive picture can qualitatively explain the major experimental observations. However, the picture does not apply if the system is treated as a vertical junction where the current flows in the out-of-plane direction. The SDE induced by Rashba SOC is proportional to $\overrightarrow{E} \times \overrightarrow{p}$ where $\overrightarrow{E}$ is the interfacial electric field and $\overrightarrow{p}$ is the momentum associated with the current flow. For a vertical junction, both vectors are in the out-of-plane direction and the SDE must vanish. On the other hand, the Ising SOC cannot generate finite-momentum pairing when $B_{\perp}$ and $I$ are collinear, and therefore cannot lead to $B_{\perp}$-induced SDE. 


Instead, we attribute the observed SDE to the existence of an in-plane component of the current. To make the modeling computationally affordable, we adopted a simplified model where an inclined current passed through the interface between two infinitely large monolayers, as shown in Fig.\ref{5}a. The upper layer is equipped with an Ising SOC of strength $+\beta$ and a Rashba SOC of $+\alpha$, the lower layer has an Ising SOC of $-\beta'$ and a Rashba SOC of $+\alpha'$, where $\beta \gg\alpha$. In the presence of a phase difference $\varphi $ between the superconductivity order parameters of the two layers, a Josephson current flows between them, which forms the out-of-plane current component $I_o$, from which we obtain the final critical Josephson currents $I_{c\pm }$. An in-plane component is taken into account by assuming an in-plane Cooper pair momentum $q$. When $q=0$ and the current is normal to the plane, the SDE vanishes for reasons discussed previously. On the other hand, our calculation shows that the SDE is nonzero for both $B_{\perp}$ and $B_{||}$ for purely in-plane supercurrent, i.e., when $\varphi =0$ (see SI for more information). With this being said, the out-of-plane $I_o$ and in-plane current $I_p$ contribute  to $\eta=\frac{I_{c+}-|I_{c-}|}{I_{c+}+|I_{c-}|}$ in a different manner. $I_p$ dominates $I_{c+}-|I_{c-}|$ term, $I_o$ primarily affects $I_{c+}+|I_{c-}|$ term. For nonzero $q$ and $\varphi $, the calculated SDE efficiency $\eta$ is shown in Fig.\ref{5}b to d with different combinations of spin-orbit coupling. Although the calculated SDE strength shows a rather different functional form from the experimental data, the calculations suggest that the dual-mode SDE originates from an interplay of the Ising and Rashba SOC, which affects the spin textures on the Fermi surfaces near the $\Gamma $ point, as shown in the inset of Fig.\ref{5}b. It is particularly interesting to note that in the small field-limit, $B_{\perp}$- and $B_{||}$-induced SDE indeed have comparable maximum diode efficiency $\eta$, 9$\%$ versus 10$\%$. We also observe a sign swapping in $\eta$ in the presence of a large $B_{\perp}$ according to the model, however, the suppression of $I_c$ may hinder the validation of this specific result in the experiment. When the Ising SOC is turned off as shown in Fig.\ref{5}c, the SDE for $B_{\perp}$ vanishes and that for $B_{||}$ is decreased by an order. Note that the sign of the SDE is also changed by the presence of Ising SOC, indicating a qualitative impact of the Ising SOC with an in-plane magnetic field. When Ising SOC is preserved and Rashba SOC is turned off, as depicted in Fig.\ref{5}d,  $B_{\perp}$-induced SDE remains significant due to $\beta \neq \beta' $ ($\eta$ peaks at 6$\%$ in the small field regime), while the $B_{||}$-signal vanishes as expected. The enhancement in $\eta$ due to the coexistence of Ising and Rashba SOC is consistent with a recent theoretical work that considers the SDE due to a similar interplay in thin flakes\cite{BAL25}.  

Then, let us switch to vortex dynamics, another major candidate. In this case, it is attempting to attribute $B_{\perp}$- and $B_{||}$-induced SDE to the Abrikosov vortex in the flake and Josephson vortex across the junction, respectively. As for our results, we cannot rule out the possibility that the $B_{\perp}$-induced SDE may be linked to the Abrikosov vortex, but we are certain $B_{||}$-induced SDE is not due to Josephson vortex.  For instance, associating $B_{\perp}$-induced SDE to the Abrikosov vortex can offer a satisfactory explanation of the results of SDE peaks 2 and 3. Now that the $B_{\perp}$ component of these peaks stays the same, the dynamics of the Abrikosov vortex would also tend to stay the same, and then $\eta$ should take a constant value.  This is indeed what happens in the experiment. However, SDE arising from the Abrikosov vortex in the previous works typically results in a more linear-like\cite{GPB24, WHM24} or highly nonmonotonic\cite{ZWW23} temperature dependence of $\eta$, rather than the square-root like one observed in our experiment. The disagreement in the temperature dependence may be a result of the difference in device fabrication or disorder strength\cite{NFW23}, hence we remain cautious on the comparison between different reports. Therefore, the potential correlation between the observed $B_{\perp}$-induced SDE and Abrikosov vortex calls for further studies.  On the other hand, associating $B_{||}$-induced SDE with the Josephson vortex is not consistent with the results of SDE peaks 1 and 4. As already mentioned, in the Josephson vortex scenario, one would expect that $B_{pos}$ of $B_{||}$-induced SDE peaks 1 and 4 to be temperature sensitive which is contrary to the experiment results.   

Apart from the underlying mechanisms, it is exciting to highlight the potential of the dual-mode SDE in realizing advanced functionality as well.  Here, we propose an exemplified device scheme, as illustrated in Fig.\ref{6}, where the dual-mode SDE has two operational functionalities: 1. Fast polarity-switching functionality. It is straightforward to generate a local magnetic field on the order of $\sim$1 mT by on-chip nanomagnets\cite{LLS23, AAS23, NCR22}, as already implemented in the control of spin qubit, to enable $B_{\perp}$-induced SDE. The nanomagnets can be driven by pulse signal and achieve a switching frequency over 100 MHz\cite{NCR22}, thus potentially allowing a rapid swapping of the polarity of SDE. 2. High-fidelity functionality. Operating the device in $B_{||}$-induced SDE mode needs a magnetic field on the order of $\sim$100 mT, considerably larger than the local magnetic field and its fluctuation in an integrated circuit. Therefore, the fidelity of the diode-polarity holding or swapping operation in $B_{||}$-induced SDE mode would be almost immune to magnetic noise in the circuit. 

Also, there are several interesting aspects to be explored in the future works. First, the observed diode efficiency of 12$\%$ may be further enhanced by encapsulating the device with hBN and optimizing the thickness of the flakes. Second, summarizing $B_{||}$-induced SDE with $B_{||}$ component pointing to different azimuthal angles may be vital in decoding the underlying mechanisms, the experiment is on-going and shows encouraging results.     

\section*{Conclusion}
Stacking 2H-$\mathrm{NbSe_2}$ and 2H-$\mathrm{NbS_2}$ flakes with similar electronic properties into heterostucture, not only enables a pronounced Josephson coupling but also activates dual-mode superconducting diode effect that can be independently generated and manipulated by both out-of-plane magnetic field $B_{\perp}$ and in-plane magnetic field $B_{||}$.  The maximum diode efficiency obtained in both modes can reach $\sim12\%$ in our prototypical device. The $B_{\perp}$-induced SDE mode is operational with a tiny magnetic field on the order of $\sim$1 mT, in stark contrast to the $B_{||}$-induced SDE mode which requires a magnetic field on the order of $\sim$100 mT. The major observations can be qualitatively understood with a simplified model considering an inclined interlayer supercurrent in the presence of both Ising and Rashba spin-orbit coupling, highlighting the role of symmetry breaking along multiple orientations. The dual-mode SDE may stimulate advanced functionalities that are not feasible for its single-mode SDE counterpart. For instance, we propose to utilize the $B_{\perp}$-induced SDE mode to implement fast polarity-switching functionality and use $B_{||}$-induced SDE mode to realize high-fidelity functionality.

\clearpage

\clearpage
\noindent \textbf{Methods}\\
\noindent\textbf{Device fabrication.} The $\mathrm{NbS_2}$/$\mathrm{NbSe_2}$ heterostructure is fabricated by a modified dry-transfer technique in a nitrogen-filled glovebox to avoid contamination and oxidization. To ensure the most optimized device quality, water and oxygen are evacuated below 1 ppm. The $\mathrm{NbSe_2}$ flake is processed and transferred first, then proceeded with the $\mathrm{NbS_2}$ flake. In the end, the $\mathrm{NbS_2}$ flake serves as the top layer while the more resilient $\mathrm{NbSe_2}$ flake works as the bottom layer. The processing protocol on $\mathrm{NbSe_2}$ flakes is quite standardized, therefore we only focus on the treatment of the $\mathrm{NbS_2}$ flakes in this section. The process starts with mechanical exfoliation of $\mathrm{NbS_2}$ bulk crystals and placing the resultant flakes on top of silicone elastomer polydimethylsiloxane (PDMS) substrates. It is particularly important to stress that the PDMS substrates are cleaned with isopropyl alcohol (IPA) before transferring the $\mathrm{NbS_2}$ flakes onto it, to avoid contamination. Then flakes of a uniform surface, desired geometry, and suitable thickness are chosen and cleaned with acetone, IPA, deionized water, and oxygen plasma at 300 W for 5 min, so that it is free of resist residues. Finally, the $\mathrm{NbS_2}$ flake are stacked on the top of $\mathrm{NbSe_2}$ flake. We do not encapsulate the prototypical devices with hBN, in order to simplify the fabrication. However, we anticipate that the encapsulation can possibly result in a more optimized device performance.     

We stress again that we have specifically picked $\mathrm{NbS_2}$ and $\mathrm{NbSe_2}$ flakes with similar thickness, because these flakes would have similar electronic properties such as order parameters and thus allow a pronounced Josephson coupling between the flakes. The dimension of the flakes are summarized in table\ref{table2}.   \\
\begin{table*}[h]
\centering
\begin{tabular}{|p{20mm}|p{30mm}|p{30mm}|p{30mm}|p{20mm}|p{20mm}|} 
 \hline
device index & thickness of bottom flake & thickness of upper flake & area of the junction & $\eta_{max}$ with $B_{\perp}$ & $\eta_{max}$ with $B_{||}$\\ 
 \hline
 heterostructure device 1 & $\mathrm{NbSe_2}$, 33 nm & $\mathrm{NbS_2}$, 46 nm & 10.26 $\mu$m$^2$ & 12$\%$ & 12$\%$\\ 
 \hline
 heterostructure device 2 & $\mathrm{NbSe_2}$, 22 nm & $\mathrm{NbS_2}$, 40 nm & 16.66 $\mu$m$^2$ & 8$\%$ & 7$\%$\\ 
 \hline
 homostructure device 3 & $\mathrm{NbSe_2}$, 29 nm & $\mathrm{NbSe_2}$, 34 nm & 11.88 $\mu$m$^2$ & 0.8$\%$ & 3$\%$\\ 
 \hline
  homostructure device 4 & $\mathrm{NbSe_2}$, 30 nm & $\mathrm{NbSe_2}$, 61 nm & 48.65 $\mu$m$^2$ & 0.6$\%$ & 0.5$\%$\\ 
 \hline
\end{tabular}
\caption{Summary on the dimension of devices studied in the work.  }
\label{table2}
\end{table*}


\vspace*{10pt}
\noindent \textbf{Acknowledgments} 

\noindent We thank X. Liu for fruitful discussions. We acknowledge the support from the National Natural Science Foundation of China (12204184, 12074134).\\

\noindent \textbf{Author contributions} 

\noindent  C. Y. conceived and designed the experiment. H. G. fabricated the samples. C. Y. and H. G. conducted experiments. H. G. acquired the data. C. Y. and S. W. analyzed the data. Q. C., X. Z. and C. Z. assisted in the experiment setup. Z. Z. and Y. S. performed the theoretical calculation under the instruction of J. H.. C. Y. and S. W. proposed the two functionality device scheme. C.Y. wrote the manuscript with input from J. H., S. W. and other authors. C.Y., J. H. and S. W. supervised the project.  All authors discussed the results and contributed to the manuscript.\\

\noindent \textbf{Competing interests} 

\noindent The authors declare no competing interests. \\

\noindent \textbf{Data availability}

\noindent The data that support the findings of this study are available from the corresponding authors upon reasonable request. \\

\clearpage

\renewcommand{\figurename}{Extended Data Figure}
\clearpage
\setcounter{figure}{0} 

\begin{figure}
	\centering
	\includegraphics[width=0.8\textwidth]{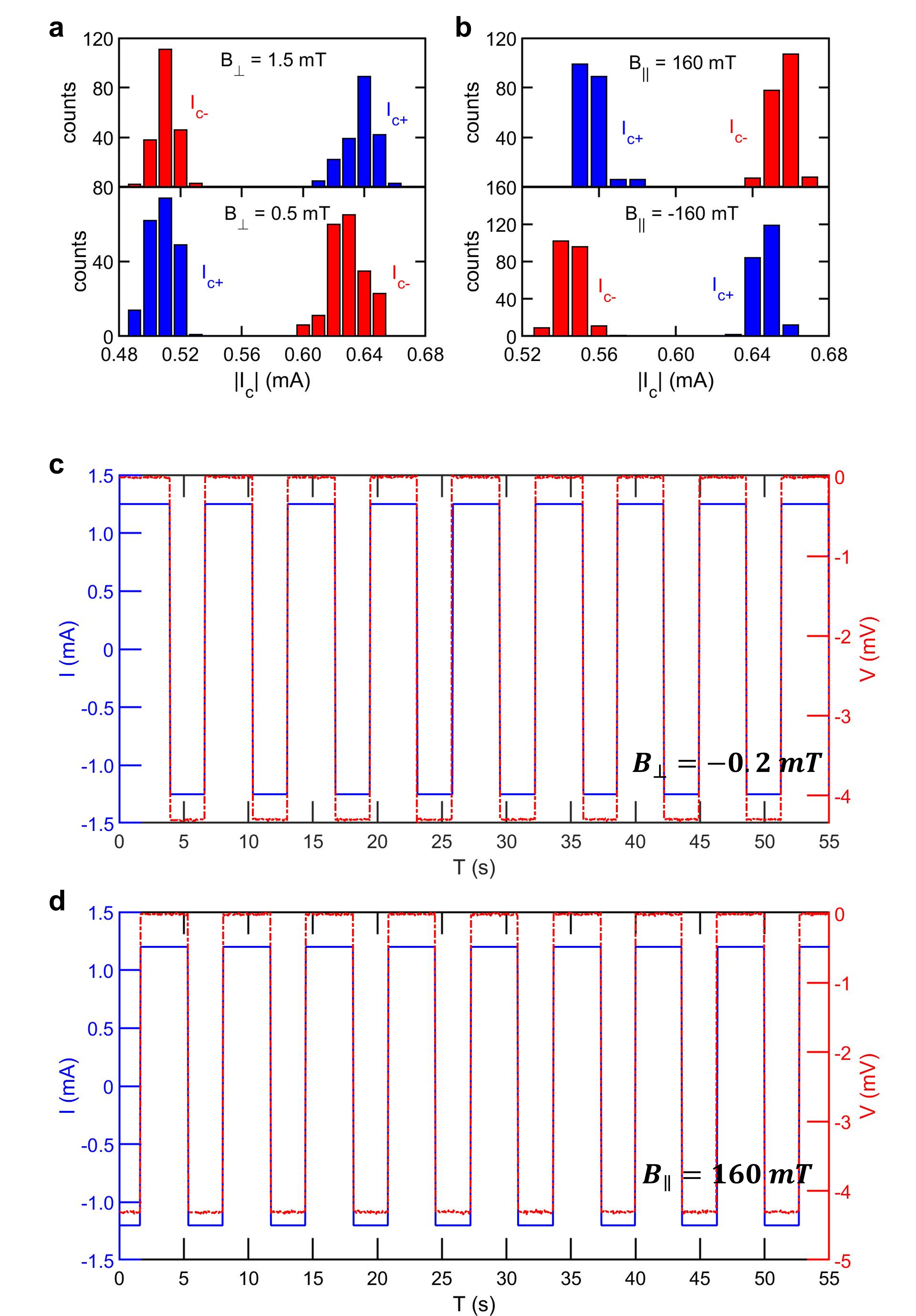}
	\caption{Statistic analysis of critical current and long-term stability of superconducting diode effect in heterostructure devices. \textbf{a\&b}, $I_{c+}$ can $I_{c-}$ can be statistically discriminated in the presence of out-of-plane and in-plane magnetic field. $B_{\perp}= 1.5$ mT and 0.5 mT correspond to the two antinodes in Fig.\ref{2}. The statistic results yield that the nonreciprocity of critical current does not arise from fluctuation of critical current. \textbf{c\&d}, half-wave rectification in device 2 exhibits long-terms stability in the presence of out-of-plane and in-plane magnetic field.           }
	\label{ExtFig1}
\end{figure}
\clearpage

\begin{figure}
	\centering
	\includegraphics[width=0.8\textwidth]{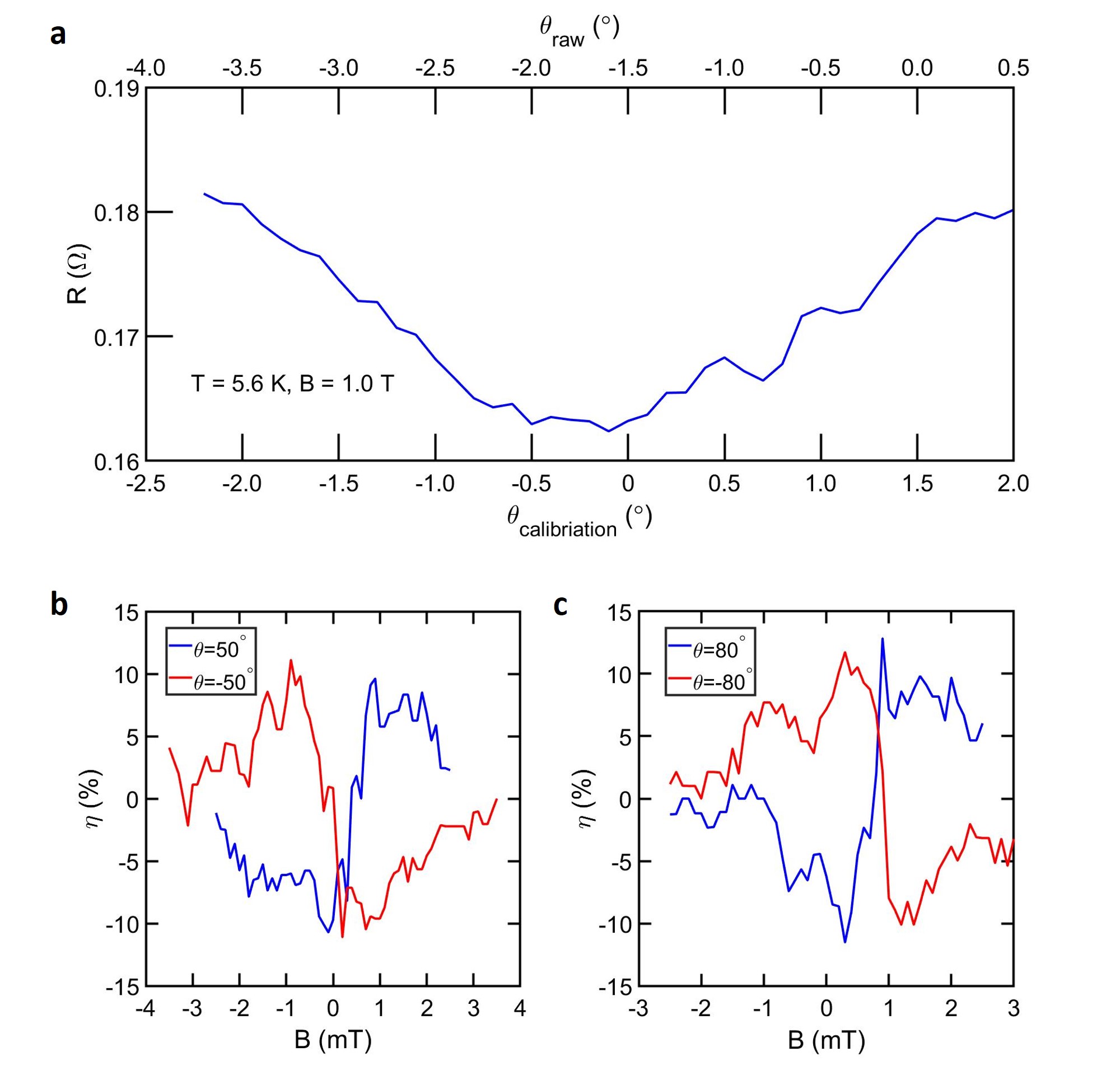}
	\caption{Representative results of superconducting diode effect at large polar angle $\theta$ in heterostructure device 1. \textbf{a}, the polar angle is calibrated by measuring the flake resistance at a function of $\theta$. $\theta$ corresponds to the minimum of flake resistance is then denoted as $0^\circ$. \textbf{b\&c}, diode efficiency as a function of magnetic field at $\theta=\pm50^\circ$ and $\pm80^\circ$, respectively.                 }
	\label{ExtFig2}
\end{figure}
\clearpage

\begin{figure}
	\centering
	\includegraphics[width=0.8\textwidth]{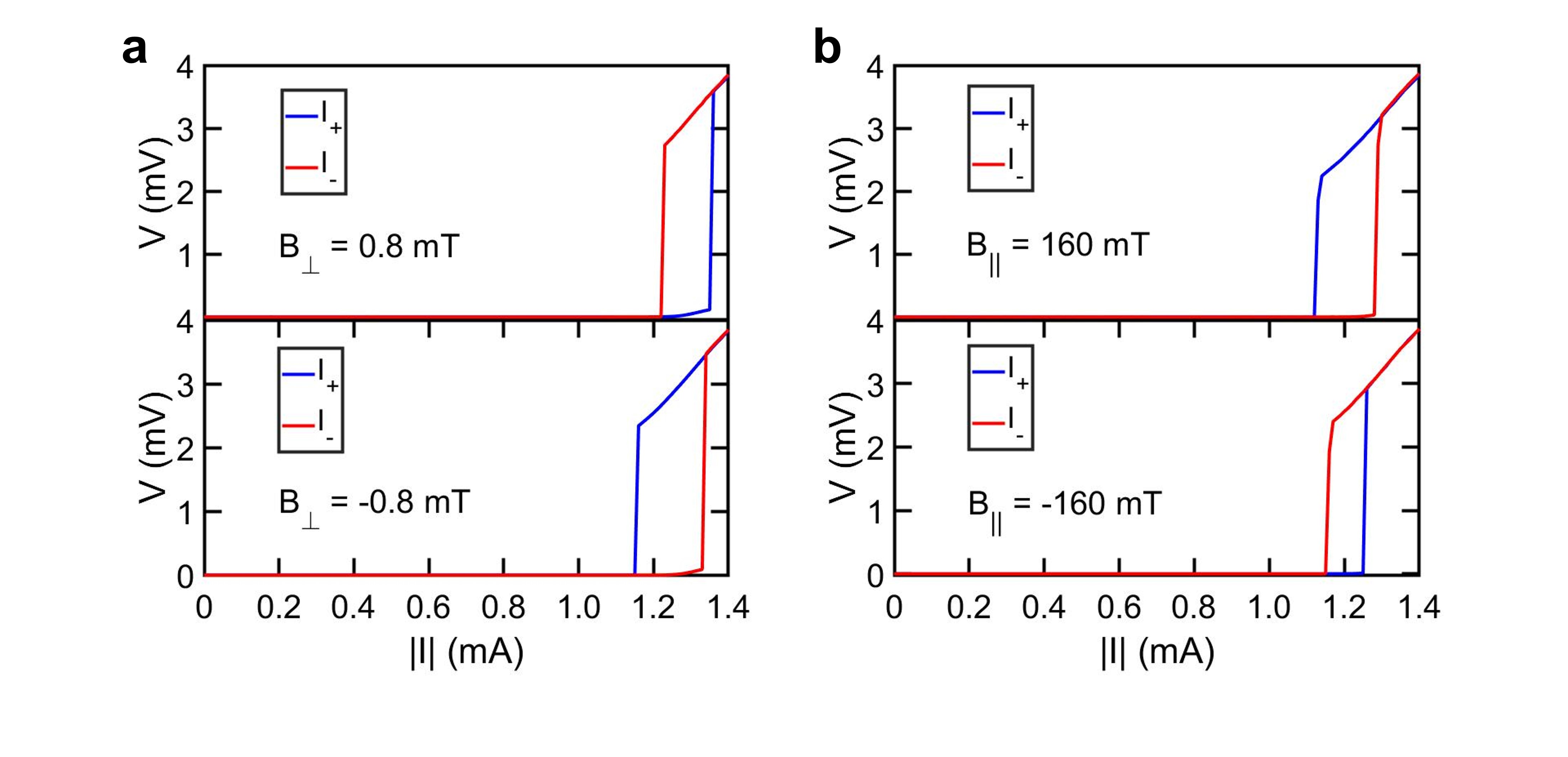}
	\caption{Nonreciprocity of critical current in heterostructure device 2. \textbf{a\&b}, representative current-voltage characteristics in the presence of out-of-plane and in-plane magnetic field. The results were recorded at the forward current sweep.             }
	\label{ExtFig3}
\end{figure}
\clearpage

\begin{figure}
	\centering
	\includegraphics[width=0.8\textwidth]{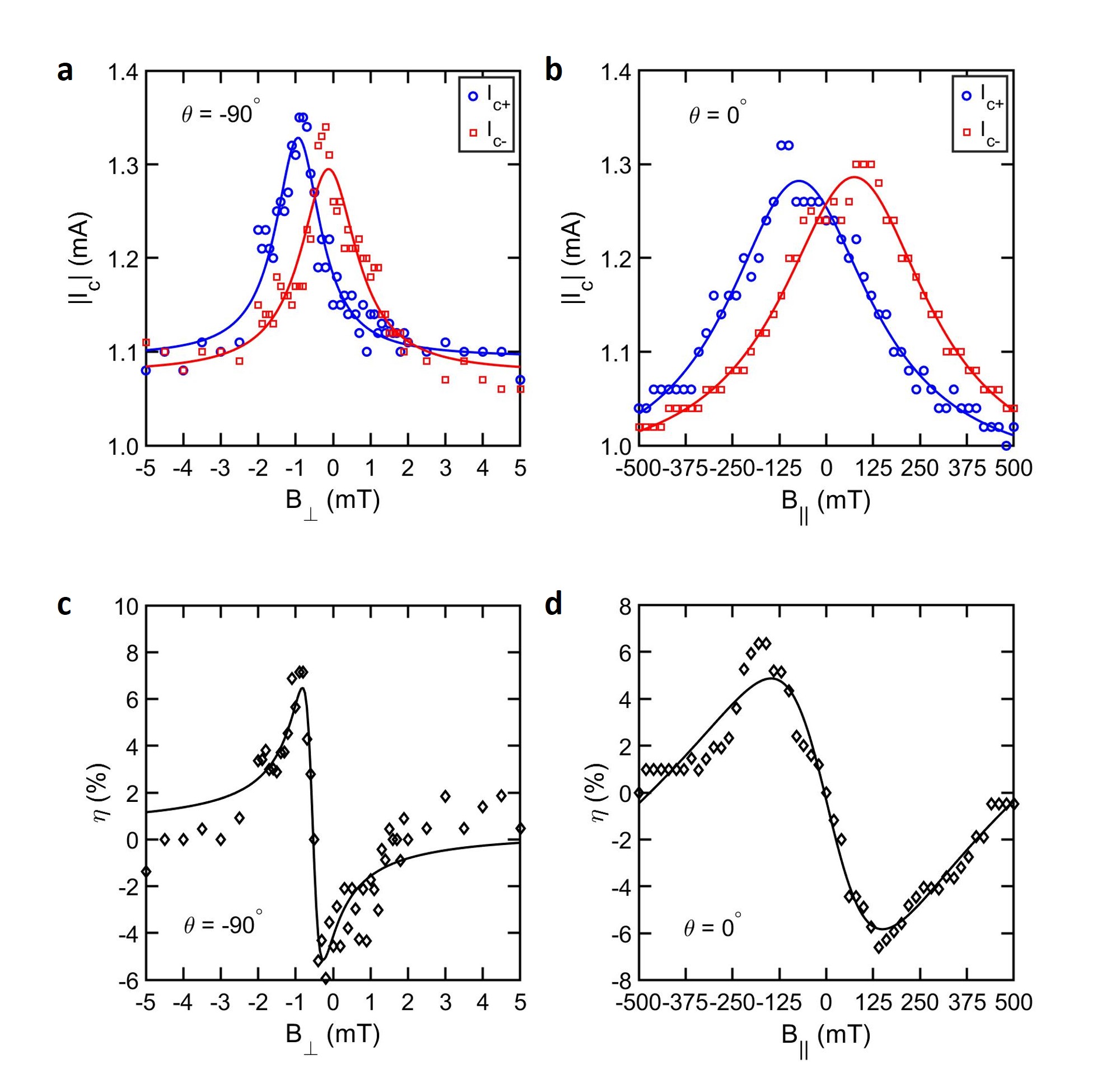}
	\caption{Evolution of superconducting diode effect as a function of magnetic field in heterostructure device 2. \textbf{a\&c}, Critical current in both current bias branches and diode efficiency $\eta$ with respect to out-of-plane magnetic field ($\theta= -90^\circ$). \textbf{b\&d}, Critical current and $\eta$ with respect to in-plane magnetic field ($\theta= 0^\circ$). The solid curves in the plots serve as guide to the eye. It is necessary to stress that the seemingly short flat segments in plot \textbf{b} is due to the large increment size in current sweeping in this particular measurement, rather than something intrinsic.             }
	\label{ExtFig4}
\end{figure}
\clearpage

\begin{figure}
	\centering
	\includegraphics[width=0.8\textwidth]{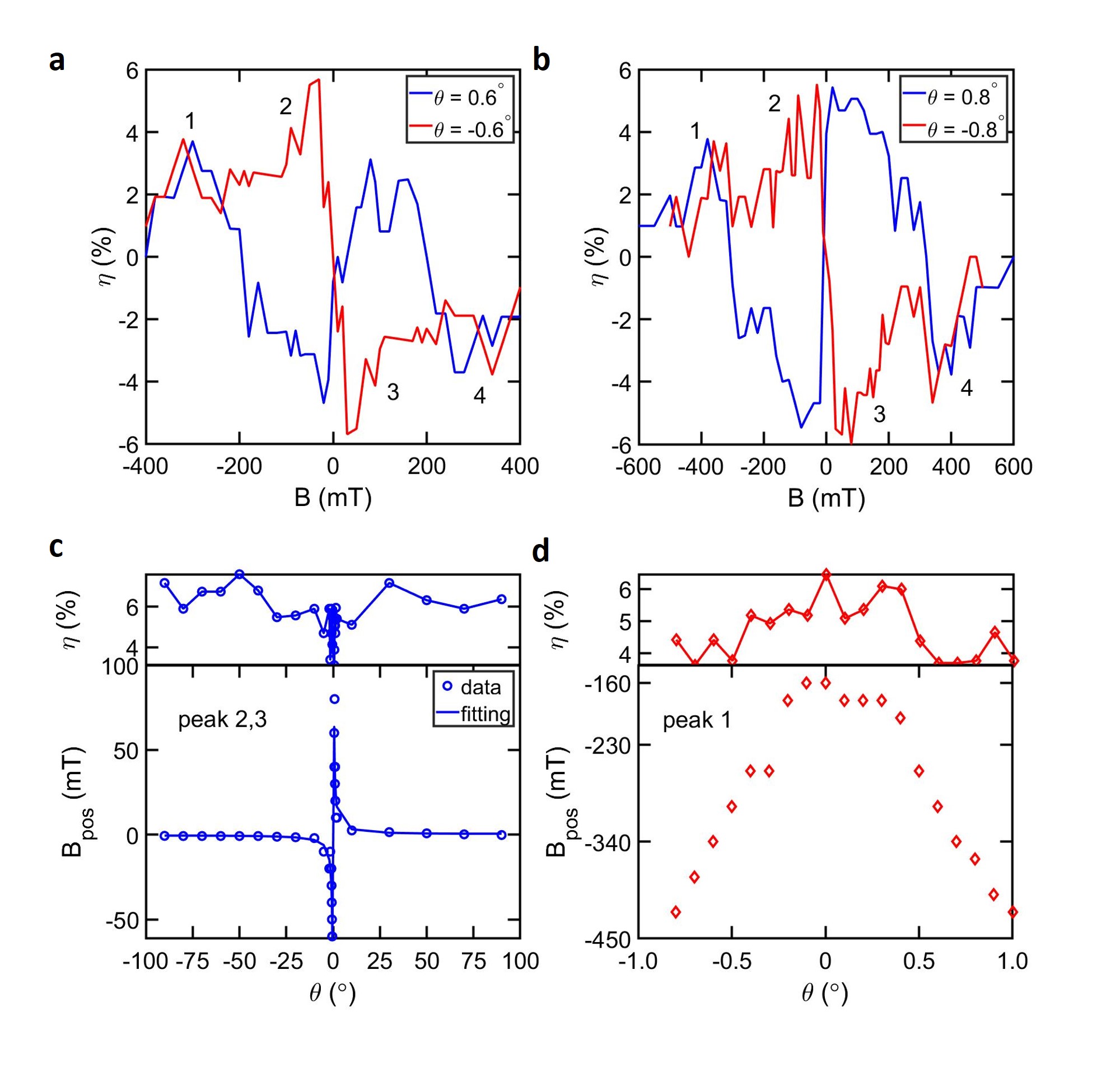}
	\caption{Superconducting diode effect at arbitrary $\theta$ in heterostructure device 2. \textbf{a\&b}, $\eta$ measured at a function of magnetic field at $\theta=\pm0.6^\circ$ and $\theta=\pm0.8^\circ$, respectively. Peaks 1 and 4 are dominated by in-plane magnetic field component $B_{||}$. peaks 2 and 3 are activated by out-of-plane magnetic field component $B_{||}$. \textbf{c}, diode efficiency $\eta$ and peak position $B_{pos}$ of $B_{\perp}$-induced SDE peaks. $B_{pos}$ follows a $1/sin(\theta)$ trend as indicated by the solid fitting curve. Again, we stick to the $\eta>0$ branch in the plot, hence $B_{pos}$ measures position of peak 3 at $\theta>0$ and peak 2 at $\theta<0$. \textbf{d}, $\eta$ and $B_{pos}$ of the $B_{||}$-induced SDE peak 1. The results are qualitatively in agreement with that obtained in device 1.                          }
	\label{ExtFig5}
\end{figure}
\clearpage

\begin{figure}
	\centering
	\includegraphics[width=0.8\textwidth]{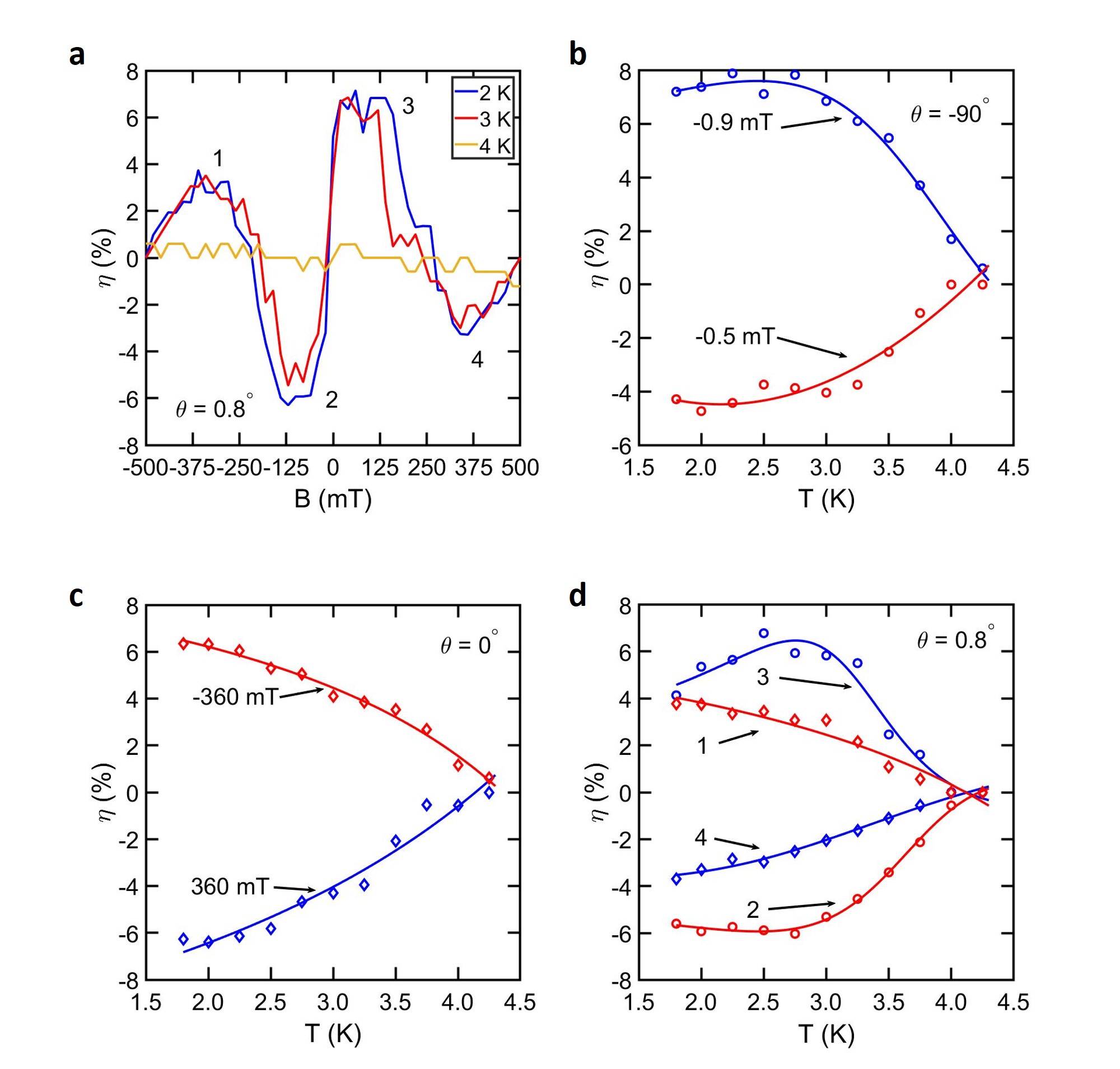}
	\caption{Temperature dependence of superconducting diode effect in heterostructure device 2. \textbf{a}, $\eta$ as a function of magnetic with increasing temperature. \textbf{b} to \textbf{d} show $\eta$ as a function of temperature at $\theta=-90^\circ$, $\theta=0^\circ$ and $\theta=0.8^\circ$. B = -0.9 mT and B = -0.5 mT in plot \textbf{b} corresponds to the two maxima in Extended Data Fig.\ref{4}c. Index 1-4 in plot \textbf{d} refer to peaks 1-4 in plot \textbf{a}. $\eta$ of $B_{\perp}$-induced SDE, plot \textbf{b} or peaks 2 and 3 in plot \textbf{d}, exhibits a highly nonlinear temperature dependence; $\eta$ of $B_{||}$-induced SDE, plot \textbf{c} or peaks 1 and 4 in plot \textbf{d}, take a more linear-like temperature dependence. These observations are consistent with that accquired in device 1. The solid curves in the plots serve as guide to the eye.         }
	\label{ExtFig6}
\end{figure}
\clearpage

\begin{figure}
	\centering
	\includegraphics[width=0.8\textwidth]{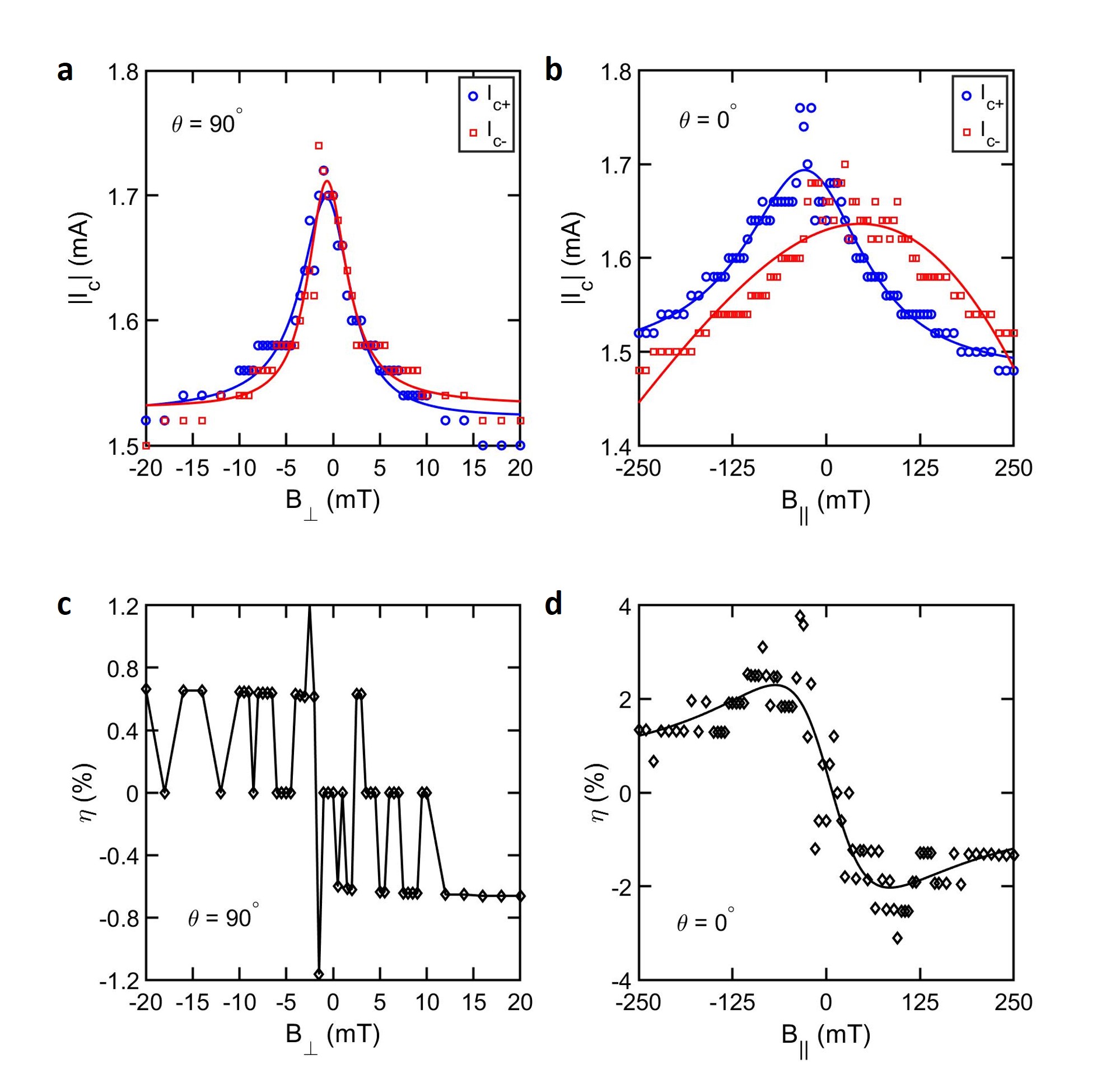}
	\caption{Nonreciprocity of critical current in $\mathrm{NbSe_2}$/$\mathrm{NbSe_2}$ homostructure device 3. \textbf{a\&c}, Critical current in both current bias branches and diode efficiency $\eta$ with respect to out-of-plane magnetic field ($\theta= -90^\circ$). \textbf{b\&d}, Critical current and $\eta$ with respect to in-plane magnetic field ($\theta= 0^\circ$). The solid curves in the plots serve as guide to the eye. $B_{\perp}$-induced SDE is almost absent in this device, while $B_{||}$-induced SDE is rather weak. In $\mathrm{NbSe_2}$/$\mathrm{NbSe_2}$ homostructure device 4, SDE is negligible regardless of the orientation of the magnetic field. }
	\label{ExtFig7}
\end{figure}
\clearpage

\end{document}